\documentclass{article}
\usepackage[T1]{fontenc}
\usepackage{caption, subcaption}
\usepackage[justification=centering]{caption}
\usepackage{adjustbox}
\usepackage{spconf}
\usepackage{mathtools,xparse}
\usepackage{amsmath,amssymb,fixmath}
\usepackage{graphicx}
\usepackage{float}
\usepackage{cite}
\usepackage[dvips]{epsfig}
\usepackage{multirow, multicol}
\usepackage{makecell}
\usepackage{multicol}
\usepackage{tabularx}

\usepackage{hhline}
\usepackage{arydshln}

\usepackage{nicefrac}

\usepackage{color}

\usepackage{verbatim}

\usepackage{tablefootnote}

\PassOptionsToPackage{hyphens}{url}\usepackage{hyperref}

\newcommand{\minjaeedit}[1]{\textcolor{black}{#1}}

\newcommand{\mosnat}{$4.56$}

\newcommand{\mosLPas}{$4.11$}
\newcommand{\mosPWGas}{$3.84$}
\newcommand{\mossrc}{$3.99$}
\newcommand{\mosFS}{$2.68$}
\newcommand{\mosFSaug}{$3.55$}
\newcommand{\mosFSmix}{$3.74$}

\newcommand{\mossrcL}{$4.22$}
\newcommand{\mosFSL}{$3.47$}
\newcommand{\mosFSaugL}{$3.80$}
\newcommand{\mosFSmixL}{$3.95$}
\newcommand{\mosT}{$2.89$}
\newcommand{\mosTaug}{$3.70$}
\newcommand{\mosTmix}{$3.72$}

\title{TTS-by-TTS: TTS-driven Data Augmentation for \\ Fast and High-Quality Speech Synthesis}

\name{Min-Jae Hwang$^{1}$, Ryuichi Yamamoto$^{2}$, Eunwoo Song$^{3}$ and Jae-Min Kim$^{3}$}
\address{$^{1}$Search Solutions Inc., Seongnam, Korea, \\ $^{2}$LINE Corp., Tokyo, Japan \\ $^{3}$NAVER Corp., Seongnam, Korea}

\begin{document}
	\maketitle
    \fontsize{9.1}{10.0}\selectfont

	\begin{abstract}
    In this paper, we propose a text-to-speech (TTS)-driven data augmentation method for improving the quality of a non-autoregressive (AR) TTS system.
    Recently proposed non-AR models, such as FastSpeech 2, have successfully achieved fast speech synthesis system.
    However, their quality is not satisfactory, especially when the amount of training data is insufficient. 
    To address this problem, we propose an effective data augmentation method using a well-designed AR TTS system. 
    In this method, large-scale synthetic corpora including text-waveform pairs with phoneme duration are generated by the AR TTS system and then used to train the target non-AR model. 
    Perceptual listening test results showed that the proposed method significantly improved the quality of the non-AR TTS system.
    In particular, we augmented five hours of a training database to 179 hours of a synthetic one. 
    Using these databases, our TTS system consisting of a FastSpeech 2 acoustic model with a Parallel WaveGAN vocoder achieved a mean opinion score of \mosFSmix, which is 40\% higher than that achieved by the conventional method.
	\end{abstract}
	\begin{keywords}
		Speech synthesis, text-to-speech, TTS-driven data augmentation, FastSpeech, Parallel WaveGAN
	\end{keywords}
	
	\section{Introduction}
	Recently proposed end-to-end text-to-speech (TTS) systems, which generate a speech signal directly from an input text, provide high-quality synthetic speech \cite{yuxuan2017tacotron, jonathan2017natural, ren2019fastspeech, ren2020fastspeech2, ping2018clarinet}.
	Popular end-to-end TTS systems consist of two subsystems: a sequence-to-sequence acoustic model, which generates the acoustic features of the speech signal from the input text, and a neural vocoder, which generates the speech waveform from the acoustic features.
    
    Two approaches have focused on the acoustic model: \textit{autoregressive} (AR) and \textit{non-AR} approaches. 
    In AR approach-based models, including Tacotron, the acoustic features are sequentially generated by conditioning previously generated ones \cite{yuxuan2017tacotron, jonathan2017natural}.
    As the models efficiently learn the temporal variation of acoustic features during the training procedure, they can provide a high-quality synthetic sound.
    However, the synthesis speed is slow due to the nature of sequential generation. 
    In contrast, non-AR approach-based models, such as FastSpeech, can generate acoustic features in parallel \cite{ren2019fastspeech, ren2020fastspeech2}.
    Thus, their generation speed is significantly faster than that of AR models and more suitable for real-time TTS applications. 
    However, due to the limited capacity of non-AR modeling, there is room for improvement of their synthesis quality, especially when the training database is not sufficient.
    
    To improve the quality of non-AR TTS, we propose a TTS-driven data augmentation method. 
    In this system, the database for training \textbf{\textit{target}} non-AR TTS (i.e., text-waveform pairs with phoneme duration) is generated by a well-designed \textbf{\textit{source}} AR TTS system. 
    First, we collect a large amount of text scripts while maintaining the recording script's phoneme distribution.
    Second, the Tacotron 2-based acoustic model generates acoustic features and phoneme durations from the collected texts. In detail, we adopted Tacotron 2 with a duration predictor \cite{okamoto2019tacotron} because it has the capacity to accurately match the alignment between phonemes and acoustic features.
    Finally, a neural excitation vocoder synthesizes the speech waveforms from the generated features.
    Among the various types of vocoders, we chose an LP-WaveNet vocoder due to its good quality with stable generation \cite{hwang2018lp}.
    After generating large-scale synthetic TTS corpora, these are used to train the target TTS system.
    As a large amount of text scripts enables the model to simulate various phoneme combinations, the target model’s stability to the unseen text can be significantly improved.
    
    We evaluated the proposed method via subjective listening tests. 
    Specifically, the target non-AR TTS systems consisting of the FastSpeech 2 acoustic model \cite{ren2020fastspeech2} with a Parallel WaveGAN vocoder \cite{yamamoto2020parallel} were trained by five hours of recorded data and 179 hours of augmented data. 
    Consequently, our system achieved a mean opinion score (MOS) of \mosFSmix, which is 40\% higher than that of systems trained without augmented data.

	\section{Relationship to prior work}
	As the quality of recent TTS systems has reached a natural level, several attempts have been made to apply TTS-synthesized speech databases to speech applications.
    For instance, Laptev et al. \cite{laptev2020you} and Jia et al. \cite{jia2019leveraging} improved the performance of automatic speech recognition and speech translation systems by training models with synthetic speech databases generated by Tacotron.
    
    In TTS applications, FastSpeech \cite{ren2019fastspeech} adopted the idea of using the generated output from AR model to train the non-AR model.
    Even though this and our methods commonly transfer the quality of AR model to the non-AR model, these are clearly difference:
    Our method uses the AR TTS model to increase the size of training database for \textit{data augmentation} purpose; whereas the FastSpeech uses it to re-generate training set's acoustic parameters for the purpose of \textit{knowledge distillation} \cite{hinton2015distilling}.
    To the best of our knowledge, our method is first approach to applying a TTS-driven data augmentation method to the TTS system again.

	\section{TTS-driven Data Augmentation}
	\label{sec:proposed}
	\begin{figure}[t!] 
        \centering
        \includegraphics[width=0.84\linewidth]{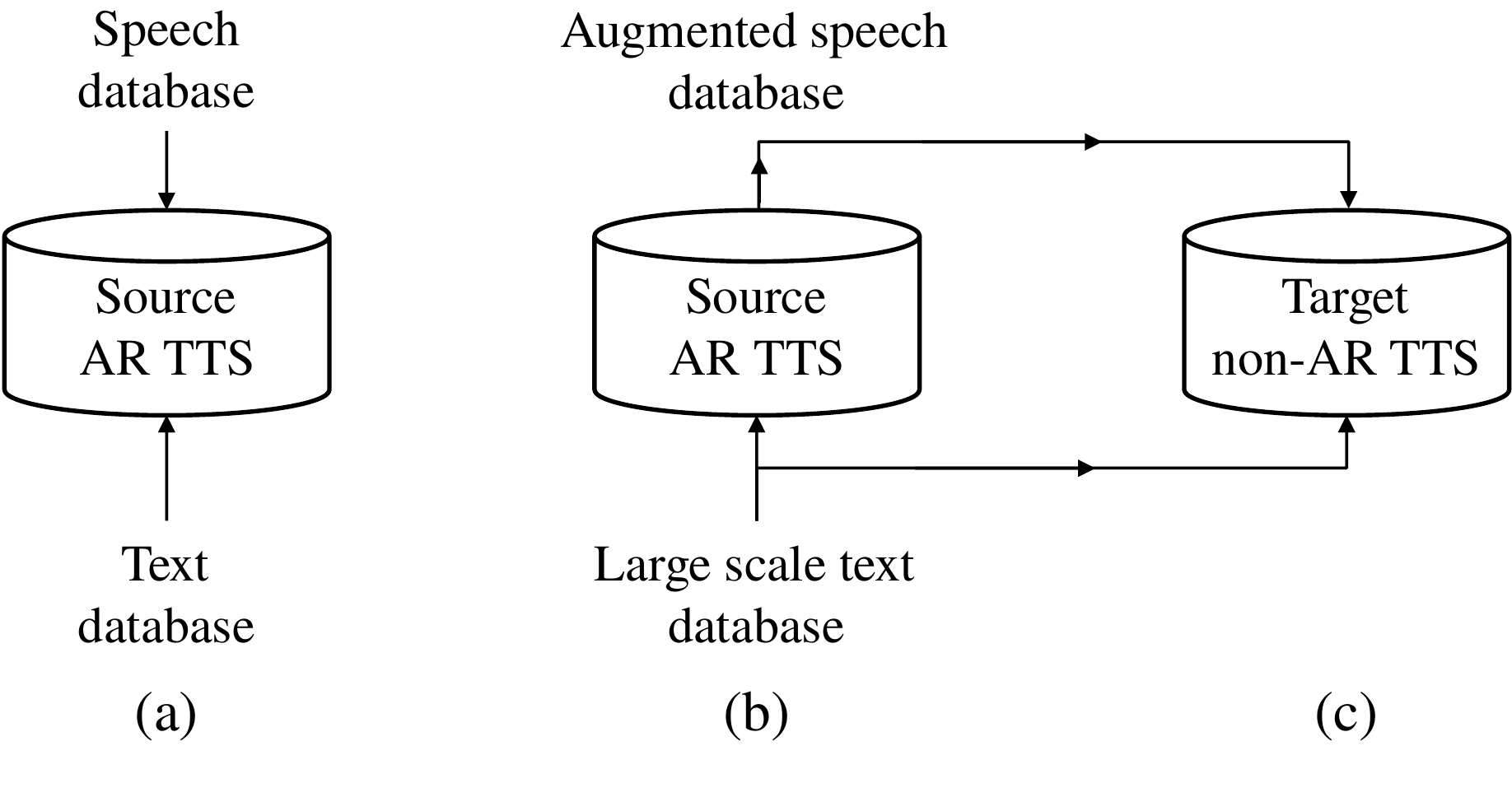}
        \vspace{-4mm}
        \caption{
            \small
    		The proposed training process with data augmentation: (a) source TTS training, (b) data augmentation, and (c) target TTS training.
        }
        \vspace{-5mm}
    	\label{fig:db_aug}
    \end{figure}

    \begin{table}[t!]
        \centering
        \caption{
            \small
            Summary of TTS systems.
        }
        \vspace{2mm}
        \label{table:tts_models_summary}
        \small
    	\setlength\tabcolsep{5.5pt}
        
        \begin{tabular}{ccc}
            \Xhline{2\arrayrulewidth}
            System    & \begin{tabular}[c]{@{}c@{}}Acoustic model\end{tabular}                    & \begin{tabular}[c]{@{}c@{}}Neural vocoder\end{tabular}  \\ \Xhline{2\arrayrulewidth} 
            \begin{tabular}[c]{@{}c@{}}Source \\ AR TTS\end{tabular}   & \begin{tabular}[c]{@{}c@{}}Tacotron 2 with\\ duration predictor \cite{okamoto2019tacotron} \end{tabular} & LP-WaveNet \cite{hwang2018lp}  \\ \hline
            \begin{tabular}[c]{@{}c@{}}Target \\ non-AR TTS\end{tabular}   & FastSpeech 2 \cite{ren2020fastspeech2}                                                  & \begin{tabular}[c]{@{}c@{}}Parallel WaveGAN \cite{yamamoto2020parallel} \end{tabular}  \\ \Xhline{2\arrayrulewidth}
            \end{tabular}
    \end{table}

    As illustrated in Fig.~\ref{fig:db_aug}, the training framework of the proposed system consists of three processes.
    First, a well-designed AR TTS model is trained by the recorded data (Fig.~\ref{fig:db_aug}-(a)).
    Then, the synthetic speech corpus with phoneme duration is generated by feeding collected text scripts to the source TTS system (Fig.~\ref{fig:db_aug}-(b)).
    Finally, the target non-AR TTS model is trained using the augmented data (Fig.~\ref{fig:db_aug}-(c)).

    \subsection{Source AR TTS model}
    \label{sec:src_ar_tts_model}
    The source and target TTS systems used for data augmentation experiments are summarized in Table.~\ref{table:tts_models_summary}.
    To generate a \textit{high-quality} synthetic TTS database, it is important to ensure that the speech generated by the source TTS model is aligned with the phonemic pronunciation. 
    Thus, we adopted a Tacotron 2 decoder with a phoneme alignment approach \cite{okamoto2019tacotron}, which has the capacity to accurately align \minjaeedit{the phoneme sequence with the} acoustic features, as an acoustic model.
    In this method, an external duration model predicts the phoneme duration from the linguistic features, and the Tacotron 2 decoder generates the corresponding acoustic features.
    Then, the LP-WaveNet-based neural excitation vocoder \cite{hwang2018lp} synthesizes the speech signals from these acoustic features.
    In this vocoder, the speech waveform is generated by the WaveNet-based mixture density network \cite{Bishop94mixturedensity} within the framework of the human speech production mechanism \cite{quatieri2006discrete}.
    As a result, it can stably generate more accurate speech signals than plain WaveNet models \cite{tamamori2017speaker, aaron2016waveNet}.

    \subsection{Data augmentation}
    \label{sec:data_augmentation}
\begin{figure}[!t]
        \centering
		\includegraphics[width=0.9\linewidth]{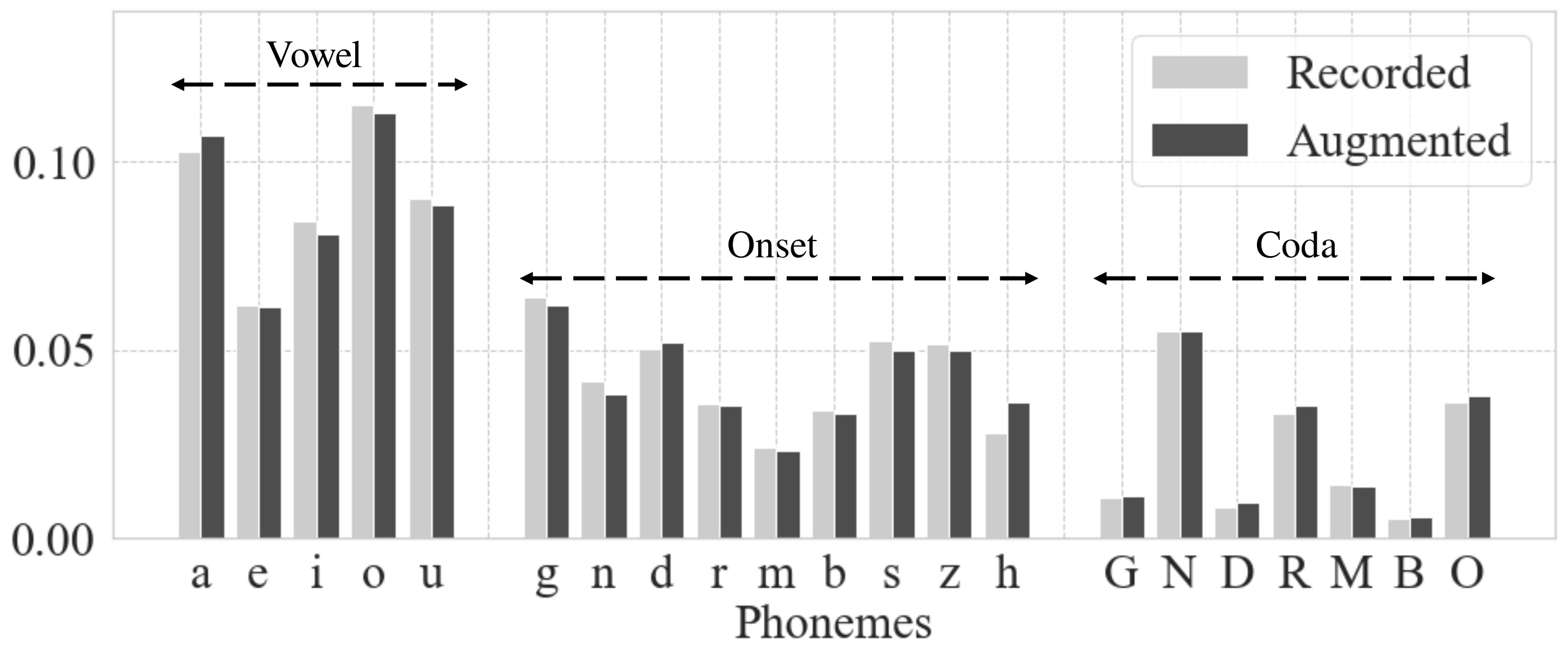}
    	\caption{\small
            Normalized histograms of phoneme distributions obtained from the recorded and augmented TTS databases.
            The numbers of phonemes used in the recorded and augmented databases were 155,715 and 6,288,422, respectively
            Note that lowercase and uppercase letters denote onset and coda consonants defined as Korean pronunciation, respectively. 
        }
        \vspace{-0mm}
    	\label{fig:phoneme_distribution}
    \end{figure}

    To prepare text scripts for data augmentation, we crawled 124,134 text scripts from news articles on the NAVER website\footnote{\url{https://news.naver.com}}.
    As shown in Fig.~\ref{fig:phoneme_distribution}, a total of 6,288,422 phonemes were collected, which was 40 times larger than the recorded database. 
    Assuming that the recorded database had a balanced phoneme distribution, we tried to crawl text scripts to follow its phoneme distribution. 
    Because a large number of phoneme sets enables the TTS model to learn various phoneme combinations, the target TTS can generate more stable synthetic speech in the condition of unseen text.

    \subsection{Target non-AR TTS model}
    \label{sec:tgt_nonar_tts_model}
    As the acoustic model of the target TTS, we adopt the state-of-the-art FastSpeech 2 model thanks to its fast inference speed and good quality \cite{ren2020fastspeech2}.
    There are several differences to its original version in our implementation.
    First, instead of using forced alignment to predict the phoneme duration \cite{mcauliffe2017montreal}, we use the phoneme duration used for source TTS system because it is already matched with the synthetic speech waveform.
    Second, \minjaeedit{to avoid the synthetic artifacts as reported in FastPitch \cite{lancucki2020fastpitch},} the pitch and energy \minjaeedit{modeled} in the variance adaptor are averaged over every input symbol by using the given durations.
    Finally, the PostNet module of Tacotron 2 \cite{jonathan2017natural} is used to improve the generation accuracy of acoustic features.
    
    To synthesize the speech waveform from the generated acoustic features, we use the Parallel WaveGAN vocoder \cite{yamamoto2020parallel}.
    This is a non-causal WaveNet model that generates a speech waveform within a generative adversarial network framework \cite{ian2014generative}.
    As adversarial training enables realistic waveform generation, the WaveNet model can efficiently generate a speech signal of good quality faster than real time. 
    The detailed configurations of the source and target TTS systems are described in Sec.~\ref{sec:model_configuration}.

	\section{Experiments}
	\subsection{Speech database}
	To train the source TTS model, a phonetically and prosodically balanced TTS corpus recorded by a female Korean professional speaker was used.
    The speech signals were sampled at 24 kHz with 16-bit quantization. 
    In total, 2,970 utterances (five hours), 590 utterances (one hour), and 290 utterances (30 minutes) were used for the training, validation, and testing sets, respectively.
    
    As described in Sec.~\ref{sec:data_augmentation}, the augmented TTS corpus was used to train the target TTS model. 
    In total, 118,734 synthetic utterances (179 hours) and 5,400 synthetic utterances (eight hours) were used for the training and validation sets, respectively.

    \subsection{Model configuration}
    \label{sec:model_configuration}
    In all model training, the input and output features were normalized to have zero mean and unit variance. 
    The weights were first initialized by the \textit{Xavier} initializer \cite{xavier2010init}, and then trained using an \textit{Adam} optimizer \cite{diederik2014adam}.
    Neural vocoders trained by the recorded database only were used in all experiments\footnote{
    It has been reported that using large size of training data is not crucial for neural vocoder \cite{okamoto2020realtime}.
    In addition, in our preliminary experiments, we could not confirm quality improvements when the neural vocoder is trained by augmented database.}.

    \subsubsection{Source AR TTS system}
    In the source TTS system, of the improved time-frequency trajectory excitation vocoder were extracted every 5 ms \cite{song2017effective}, which included 40-dimensional line spectral frequencies, fundamental frequency, energy, voicing flag, 32-dimensional slowly evolving waveform, and 4-dimensional rapidly evolving waveform, all of which composed a total 79-dimensional feature vector.
    
    The acoustic model of the source TTS consists of three sub-modules: a context analyzer, a context encoder, and a Tacotron decoder.
    In the context analyzer, 354-dimensional phoneme-level linguistic feature vectors consisting of 330 categorical and 24 numerical contexts were first extracted from the input text. 
    Then, the duration predictor, which consists of three fully connected (FC) layers with 1,024, 512, and 256 units, and a long short-term memory (LSTM) layer with 128 memory blocks, estimated the duration of each phoneme. 
    Based on the estimated duration, the phoneme-level linguistic features were upsampled to that of the frame level.
    In the context encoder, high-level context features were further extracted by feeding the frame-level linguistic features to the three convolution layers with 10$\times$1 kernels and 512 channels, bidirectional LSTM with 512 memory blocks, and FC layers with 512 units.
    Then, the Tacotron decoder, which consists of PreNet, PostNet, and main unidirectional LSTM, generated the acoustic features. 
    First, the previously generated acoustic features were fed into PreNet, which consists of two FC layers with 256 units. 
    Then, the outputs of PreNet and a context-embedding module were passed through two unidirectional LSTM layers with 1,024 memory blocks, followed by two projection layers with 79 units to generate the acoustic features. 
    Finally, PostNet, which consists of five convolution layers with 5$\times$1 kernels and 512 channels, was used to add the residual elements of the generated acoustic features for more accurate generation.
    
    In the configuration of LP-WaveNet, the dilations were set to [$2^0, 2^1, ..., 2^9$] and repeated three times, resulting in 30 layers of residual blocks and 3,071 samples of the receptive field. 
    In each residual block, 128 channels of convolution layers were used.
    The number of output dimensions was set to two to generate the mean and standard deviation of Gaussian distribution. 
    The weight normalization technique, which normalizes the weight vectors to have a unit length, was applied \cite{tim2016weight}.
    
    To improve the spectral clarity of the synthesized speech, the spectral domain sharpening filter with a coefficient of 0.95 \cite{song2017effective} was applied as a post-processing technique. 
    In addition, to generate a cleaner speech sound, LP-WaveNet’s generated scale parameter on the voiced region was reduced by a factor of 0.85 \cite{hwang2018lp}.
    
    \subsubsection{Target non-AR TTS system}
    \label{sec:config_target_noar_tts}
    In the target TTS system, an 80-dimensional Mel-spectrogram extracted every 10 ms was used as acoustic features \cite{jonathan2017natural}. 
    Like the source TTS, the acoustic model of the target TTS consisted of three sub-modules: a feed-forward Transformer (FFT) encoder, a variance adaptor, and an FFT decoder \cite{ren2020fastspeech2}.
    
    First, the phoneme sequence defined by 55 vocabulary passed through a 256-dimensional embedding layer. 
    Four FFT blocks were used in both the encoder and the decoder.
    In each FFT block, the hidden size of the self-attention layer and the number of attention heads were 384 and 2, respectively. 
    The kernel sizes of convolution layer in the two-layer convolutional network after the self-attention layer were set to 9 and 1, with input/output sizes of 384/1,024 for the first layer and 1,024/384 for the second layer.
    In the decoder, the output FC layer converted the 256-dimensional hidden states into 80-dimensional Mel-spectrograms with residual components predicted by PostNet.
    The variance adaptor consisted of three variance predictors estimating duration, pitch, and energy components, respectively.
    The variance predictor was composed of five convolution blocks, each containing 1D convolution and rectified linear unit (ReLU) activation, followed by layer normalization and dropout with a probability of 0.5. 
    The number of dimensions and the kernel size of convolution layer were set to 256 and 5, respectively. 
    The final FC layer converted 256-dimensional hidden features to the output variance parameter.

    Similar to LP-WaveNet, Parallel WaveGAN consisted of 30 layers of dilated residual convolution blocks with three dilation cycles. 
    The numbers of residual and skip channels were set to 64, and the convolution filter size was set to five. 
    The resulting receptive field of the model was 12,277.
    Weight normalization was applied to all convolutional layers \cite{tim2016weight}.
    The discriminator configuration was the same as that of Parallel WaveGAN \cite{yamamoto2020parallel}.

    \begin{figure}[t]
        \centering
        \includegraphics[width=0.85\linewidth]{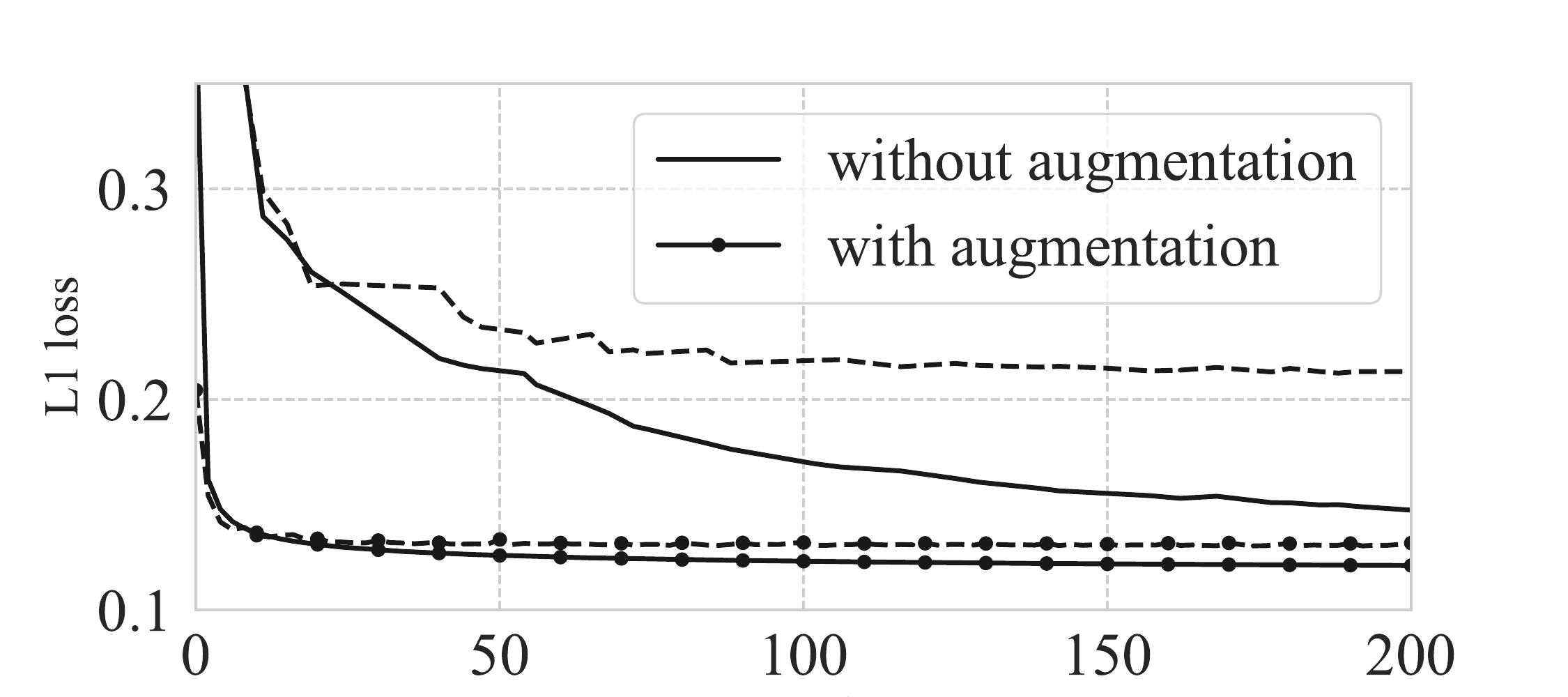}
        \caption{
            \small
            L1 loss obtained during the training process of the FastSpeech 2 model with and without augmentation. The solid and dashed lines show the training and validation losses, respectively. 
    	    }
        \label{fig:loss_curve}
        \vspace{-2mm}
    \end{figure}


    \subsection{Results}
    \subsubsection{Training efficiency}
    Fig.~\ref{fig:loss_curve} shows the training and validation losses obtained during the training process of the FastSpeech 2 model. 
    Our observations are summarized as follows:
    (1) with data augmentation, the model exhibited significantly less loss. This indicated that data augmentation was beneficial to accurately estimate the acoustic features. 
    (2) The gap between the training and validation losses in the augmentation case was significantly narrower than the case without augmentation.
    This indicated that model’s generalization performance was also improved where it provided the consistent estimation results to both the seen (train) and unseen (validation) data.
    \begin{table}[!t]
        \centering
        \small
        \caption{
            \small
            Subjective MOS test results with 95\% confidence intervals.
            Note that the score of the analysis/synthesis system can be considered the upper bound of the TTS system
        }
        \setlength\tabcolsep{3pt}
        \setlength\tabcolsep{2.5pt}
        \vspace*{2mm}
        \hspace*{-1mm}
        \label{table:subj_05hrs}
        \begin{tabular}{cccccc}
        \Xhline{2\arrayrulewidth}
        \multirow{3}{*}{System} & \multirow{3}{*}{Model}      & \multirow{3}{*}{\begin{tabular}[c]{@{}c@{}}Analysis / \\ Synthesis\end{tabular}} & \multicolumn{2}{c}{\multirow{2}{*}{\begin{tabular}[c]{@{}c@{}}Training\\ Database\end{tabular}}} & \multirow{3}{*}{MOS}   \\
        &&&& \\ \cline{4-5}
                                    &                              &                                       & Recorded          & Augmented          &                        \\\Xhline{2\arrayrulewidth}
        R1                    & Recorded                           & --                                    & --                & --                 & \mosnat$\pm$0.13          \\ \hline
        B1                  & \multirow{2}{*}{TTS$_{AR}$}  & Yes                                   & --               & --                 & \mosLPas$\pm$0.17         \\ 
        B2                  &                              & --                                    & Yes               & --                 & \mossrc$\pm$0.16          \\ \hline
        B3                  & \multirow{4}{*}{TTS$_{NAR}$} & Yes                                   & --               & --                 & \mosPWGas$\pm$0.16         \\ 
        B4                  &                              & --                                    & Yes               & --                 & \mosFS$\pm$0.34          \\ 
        P1                  &                              & --                                    & --                & Yes                & \mosFSaug$\pm$0.25          \\ 
        \textbf{P2}         &                              & --                                    & \textbf{Yes}               & \textbf{Yes}                & \textbf{3.74$\pm$0.20} \\ \Xhline{2\arrayrulewidth}
        \end{tabular}
        \begin{tabular}{l}
              \begin{minipage}{\linewidth}
                \scriptsize
                \vspace*{-0.cm}
                ~\\ R: recording; B: baseline; P: proposed model; TTS$_{AR}$: source AR TTS system; TTS$_{NAR}$: target non-AR TTS system.
                The non-AR TTS system with the highest score is shown in boldface. 
              \end{minipage}
        \end{tabular}
    \label{tab:my-table}
    \end{table}

    \subsubsection{Subjective listening tests}
    \label{sec:result:subjective_eval}
    To evaluate the perceptual quality of the proposed system, MOS listening tests were performed\footnote{Generated audio samples are available at the following URL:	\\ \url{https://min-jae.github.io/icassp2021/}}.
    Eighteen native Korean listeners were asked to score the randomly selected 20 synthesized utterances from the test set using a following possible 5-point MOS responses: 1 = Bad, 2 = Poor, 3 = Fair, 4 = Good, 5 = Excellent.

    \begin{table}[t!]
        \centering
        \small
        \caption{
            \small
            Subjective MOS test results with 95\% confidence intervals of augmentation applied to the end-to-end Tacotron 2 model instead of the FastSpeech 2 model. 
        }
        \vspace{2mm}
        \renewcommand{\arraystretch}{0.9}

        \label{table:subj_taco2}
        \begin{tabular}{ccccc}
        \Xhline{2\arrayrulewidth}
        \multirow{2}{*}{System} & \multirow{2}{*}{Model}      & \multicolumn{2}{c}{\begin{tabular}[c]{@{}c@{}}Training database\end{tabular}} & \multirow{2}{*}{MOS}   \\ \cline{3-4}
                               &                              & Recorded                                & Augmented                              &                        \\ \Xhline{2\arrayrulewidth}
        B4-T                     & \multirow{3}{*}{\begin{tabular}[c]{@{}c@{}}Tacotron 2\\ + PWG\end{tabular}} & Yes                                     & --                                     & \mosT$\pm$0.36          \\ 
        P1-T                     &                              & --                                      & Yes                                    & \mosTaug$\pm$0.26          \\ 
        \textbf{P2-T}            &                              & \textbf{Yes}                            & \textbf{Yes}                           & \textbf{3.72$\pm$0.32} \\ \Xhline{2\arrayrulewidth}
        \end{tabular}
        \begin{tabular}{l}
              \begin{minipage}{\linewidth}
                \scriptsize 
                ~\\B: baseline; P: proposed model; 
                T: Tacotron; 
                PWG: Parallel WaveGAN.
                The system with the highest score is shown in boldface.
              \end{minipage}
        \end{tabular}
    \end{table}

    \begin{table}[t!]
        \centering
        \small
        \caption{
            \small
            Subjective MOS test results with 95\% confidence intervals with the recorded data increased from 5 to 20 hours. 
        }
        \vspace{2mm}
        \renewcommand{\arraystretch}{0.9}
        \label{table:subj_20hrs}
        \begin{tabular}{ccccc}
        \Xhline{2\arrayrulewidth}
        \multirow{2}{*}{System} & \multirow{2}{*}{Model}      & \multicolumn{2}{c}{\begin{tabular}[c]{@{}c@{}}Training database\end{tabular}} & \multirow{2}{*}{MOS}   \\ \cline{3-4}
                               &                              & Recorded                                & Augmented                              &                        \\ \Xhline{2\arrayrulewidth}
        B2-L                     & TTS$_{AR}$                   & Yes                                     & --                                     & \mossrcL$\pm$0.15          \\ \hline
        B4-L                     & \multirow{3}{*}{TTS$_{NAR}$} & Yes                                     & --                                     & \mosFSL$\pm$0.43          \\ 
        P1-L                 &                              & --                                      & Yes                                    & \mosFSaugL$\pm$0.27          \\ 
        \textbf{P2-L}            &                              & \textbf{Yes}                            & \textbf{Yes}                           & \textbf{3.95$\pm$0.23} \\ \Xhline{2\arrayrulewidth}
        \end{tabular}
        \begin{tabular}{l}
                \begin{minipage}{\linewidth}
                    \scriptsize 
                    ~\\B: baseline; 
                    P: proposed model; 
                    L: trained by a large database; 
                    TTS$_{AR}$: source AR TTS system;
                    TTS$_{NAR}$: target non-AR TTS system.
                    The non-AR TTS system with the highest score is shown in boldface.
                  \end{minipage}
        \end{tabular}
    \end{table}

    Table~\ref{table:subj_05hrs} summarizes the MOS test results, whose trends can be analyzed as follows:
    (1) the AR TTS (TTS$_{AR}$) system performed better than the non-AR TTS (TTS$_{NAR}$) system because of the AR model's better capacity to capture the temporal variation of the speech signal.
    (2) When the non-AR TTS system was trained by the augmented database, its perceptual quality was significantly improved (B4 vs. P1).
    (3) When both the recorded and augmented databases were used to train the non-AR TTS system, its perceptual quality was further improved (P1 vs. P2).
    In particular, even though only five hours of a natural database were used, the quality of the non-AR TTS system achieved \mosFSmix{} MOS, which is 40\% higher than the system without augmentation (B4 vs. P2).

    \subsection{Additional experiments}
    \subsubsection{Data augmentation for AR TTS system}
    To examine whether the proposed method works well with attention mechanism in the end-to-end AR model, we replaced the non-AR FastSpeech 2 model with the Tacotron 2 model \cite{jonathan2017natural}. 
    The structure of Tacotron 2 was similar to our source TTS model, but it followed its original version.
    First, instead of using the external duration predictor, we used a location-sensitive attention mechanism \cite{chorowski2015attention}.
    Second, instead of using the linguistic features, the phoneme sequence was used as input.
    The detailed configuration followed the ESPnet-TTS toolkit \cite{hayashi2020espnet}. 
    Note that this architecture contained a potential alignment failures including attention skip or collapse.
	\begin{figure}[t!] 
        \centering
        \subfloat[]{%
            \includegraphics[width=0.5\linewidth]{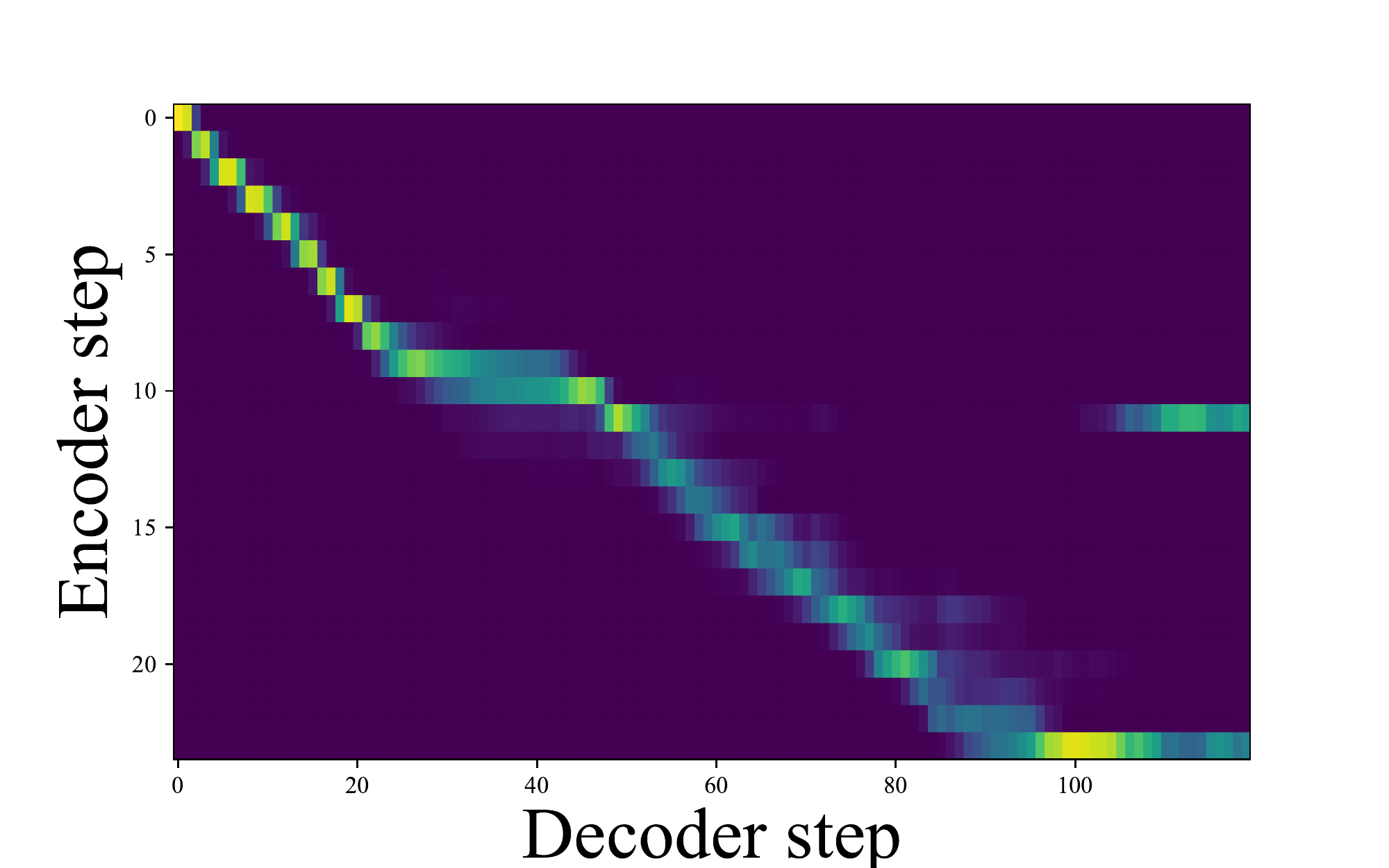}
            \label{fig:a}%
            }%
        \subfloat[]{%
            \includegraphics[width=0.5\linewidth]{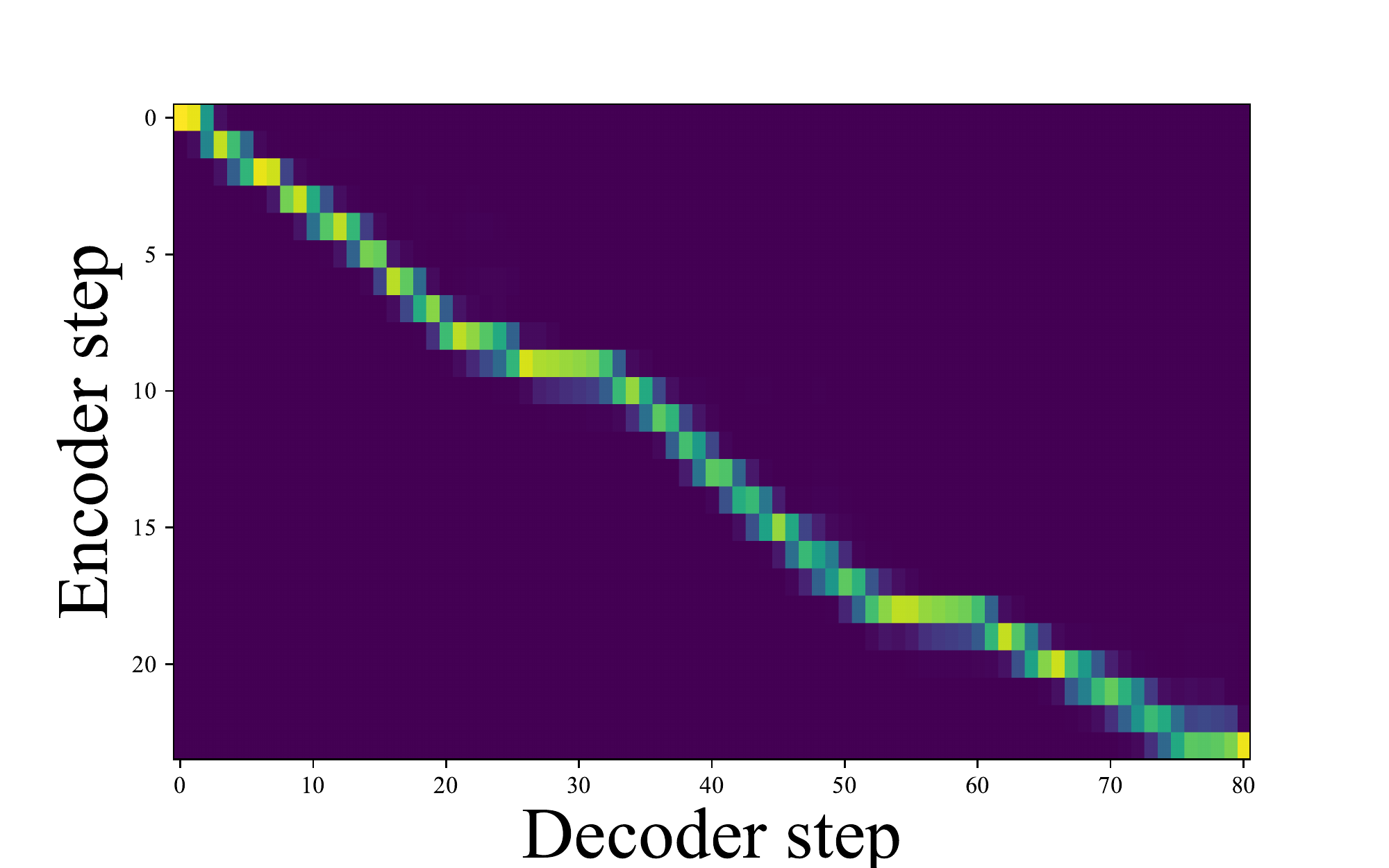}%
            \label{fig:b}%
            }%
        \vspace{-0.2cm}
        \caption{
            \small
            Attention alignments generated by Tacotron 2 acoustic models (a) without and (b) with augmentation.
        }
    	\label{fig:taco_align}
    	\vspace{-2mm}
    \end{figure}

    As shown in Fig.~\ref{fig:taco_align}, the attention alignments generated by the augmentation method were clearer than those of the model without augmentation. 
    We conjecture that the data augmentation was beneficial to improve the robustness to the unseen text patterns.
    In the subjective evaluation\footnote{\label{note1}The experimental settings for the MOS test were the same as described in Sec.~\ref{sec:result:subjective_eval}} results summarized in Table.~\ref{table:subj_taco2}, we figured out that the proposed data augmentation further improved the perceptual quality of Tacotron 2 acoustic model by achieving \mosTmix{} MOS, which is 28\% higher than the results without augmentation (B4-T vs. P1-T and P2-T).

    \subsubsection{Data augmentation with enough recordings}
    To verify the effectiveness of the proposed method with a sufficient amount of recorded data, we conducted additional experiments by increasing the size of recorded data from 2,970 (five hours) to 11,890 utterances (20 hours).
    We re-trained the source AR TTS system with the larger database and re-generated augmentation data. 
    Then, the target non-AR TTS system was also re-trained using this database. 

    The subjective evaluation\textsuperscript{\ref{note1}} results are shown in Table.~\ref{table:subj_20hrs}.
    \minjaeedit{
        The perceptual quality of source TTS was improved as \minjaeedit{the amount of training data was increased} (B2 vs. B2-L).
    }
    Moreover, the perceptual quality of the non-AR TTS system was significantly improved when the augmented database was included in the training process (B4-L vs. P1-L and P2-L).
    In particular, when both recorded and augmented database were used, the target TTS achieved \mosFSmixL{} MOS which is higher score than the case to train the model with smaller database (P2 vs. P2-L).

    \section{Conclusion}
    In this paper, we proposed a TTS-driven data augmentation method to improve a quality of non-AR TTS system.
    Using a large-scale synthetic TTS database generated by a high-quality AR TTS system, we successfully improved the quality of the target TTS system. 
    The experimental results verified that the proposed data augmentation was effective in various experimental conditions, especially when the training data were insufficient. 
    As we collected the text scripts during augmentation by keeping the recorded data's phoneme distribution, the future studies should test the augmentation with various phoneme distributions, such as uniformly distributed case.

	\section{Acknowledgements}
    This work was supported by Clova Voice, NAVER Corp., Seongnam, Korea.
    The authors would like to thank SungJun Choi and Sangkil Lee for their support.

    \vfill\pagebreak
	
    \bibliographystyle{IEEEtran}
    \label{sec:refs}
	\bibliography{mybib}

\end{document}